\documentclass[preprint,showpacs,preprintnumbers,amsmath,amssymb]{revtex4}

\usepackage{graphicx}
\usepackage{dcolumn}
\usepackage{bm}

\begin{document}

\title{Polarization and frequency disentanglement of photons via
stochastic polarization mode dispersion}

\author{Phoenix S. Y. Poon and C. K. Law}
\affiliation{Department of Physics and Institute of Theoretical
Physics, The Chinese University of Hong Kong, Shatin, Hong Kong
SAR, China}

\date{\today}
\begin{abstract}
We investigate the quantum decoherence of frequency and polarization
variables of photons via polarization mode dispersion in optical
fibers. By observing the analogy between the propagation equation of
the field and the Schr\"odinger equation, we develop a master
equation under Markovian approximation and analytically solve for
the field density matrix. We identify distinct decay behaviors for
the polarization and frequency variables for single-photon and
two-photon states. For the single photon case, purity functions
indicate that complete decoherence for each variable is possible
only for infinite fiber length. For entangled two-photon states
passing through separate fibers, entanglement associated with each
variable can be completely destroyed after characteristic finite
propagation distances. In particular, we show that frequency
disentanglement is independent of the initial polarization status.
For propagation of two photons in a common fiber, the evolution of a
polarization singlet state is addressed. We show that while complete
polarization disentanglement occurs at a finite propagation
distance, frequency entanglement could survive at any finite
distance for gaussian states.
\end{abstract}


\maketitle

\section{Introduction}
Polarization mode dispersion (PMD) in optical fibers limits fiber
performance when designing optical channels with high bit rate
\cite{PMD_review}. Physically, the origin of PMD is optical
birefringence caused by the asymmetry of the fiber due to factors
such as external mechanical stress and temperature fluctuations,
resulting in different group velocities for two orthogonal
polarization modes. In addition, stochastic optical birefringence
inside a single mode fiber leads to the random coupling of the two
polarization modes, thus causing effects such as pulse widening
\cite{PMD_review,Agrawal}, and the fluctuations of arrival times of
pulses grow with the square root of the propagation distance
\cite{Gisin1995}. Existing major applications of quantum
communication, including quantum cryptography \cite{quancryp} and
quantum teleportation \cite{teleport,teleportexpt}, rely on quantum
entanglement between photons as a crucial element. Therefore
strategies to cope with decoherence have been investigated recently,
for example, protection schemes based on decoherence-free subspace
(DFS) have been proposed \cite{DFSDuan,DFS1,DFS2,Konrad}. Typically,
a decoherence free two-photon state involves polarization singlet
states with both photons having the same frequency
\cite{singletexpt}. However, it is known that for photons with
different frequencies, decoherence by PMD cannot be avoided due to
the breaking of symmetry of the collective states
\cite{limitations}.

Entangled photon pairs produced by spontaneous down-conversion can
be hyperentangled \cite{hyperent,hyperexpt}, i.e. entangled in
various degrees of freedom (DOFs) of the photons. For example,
polarization \cite{hypershih,hyperzy}, frequency \cite{hyperfreq},
angular momentum \cite{hyperam1,hyperam2}, and energy time
\cite{energytime} are variables that can be exploited. It is thus
important to address how quantitatively hyperentangled photons
disentangle for each DOF, by investigating the decoherence of each
DOF separately. In this paper, we focus on polarization and
frequency hyperentangled photons and explore how the entanglement of
these two DOFs may affect each other inside optical fibers with
stochastic PMD.

The main purpose of this paper is to determine disentanglement
length scales associated with frequency and polarization variables
due to PMD decoherence, and to quantify the residual entanglement as
photons propagate. To this end we will investigate three situations
of photon propagation as depicted in Fig. 1. First we examine the
propagation of single photon states, then hyperentangled two-photon
states, in both polarization and frequency DOFs. In one case we
consider each photon passes through separate fibers, and in another
case both photons pass through the same fiber, all of length $L$.
For each of these cases, we approach the problem by employing the
master equation technique \cite{gardiner,bloch}, which is based on
an analogy between evolution of a state experiencing PMD along
propagation direction, and that of spin-half particles in a
stochastic magnetic field. By taking into consideration the
frequency-dependent coupling strengths of the photon states with
fiber birefringence, we analytically solve for the output density
matrices.

The organization of this paper is as follows. After providing the
basic equations of our PMD model and quantum states of photons in
Sec. II and III, we investigate the propagation of a single-photon
wave packet in Sec. IV. By obtaining the single-photon's purity
functions for each DOF, we characterize the loss of purity of each
DOF of the photon as it propagates. The results importantly provide
the characteristic decoherence lengths for individual photons. In
particular, we obtain the pulse width at the output, which defines
the minimum separation of well-resolved input pulses. In section V
and VI, we investigate the dynamics of frequency and polarization
disentanglement corresponding to the two-photon cases shown in Fig.
1b and 1c. Our analysis is based on Peres and Horodecki's powerful
criterion of entanglement, known as the PPT (positive partial
transposition) criterion \cite{PPTperes,PPThoro1, PPThoro}, and
quantum entanglement is quantified by the negativity of partially
transposed density matrix \cite{monotone}. For separate fiber
propagation (Sec. V), we show that frequency disentanglement is
independent of the initial polarization status. In addition, finite
length disentanglement is possible for both DOFs, each having
distinct characteristic length scales. In Section VI, we address the
disentanglement process in common fiber propagation, and
particularly, we examine a polarization entangled two-photon state
in the singlet form with Gaussian distribution of frequencies. The
results provide insights about the disentanglement processes for
states near DFS. Section VII is devoted to our conclusions.

\begin{figure}
\includegraphics [width=10 cm] {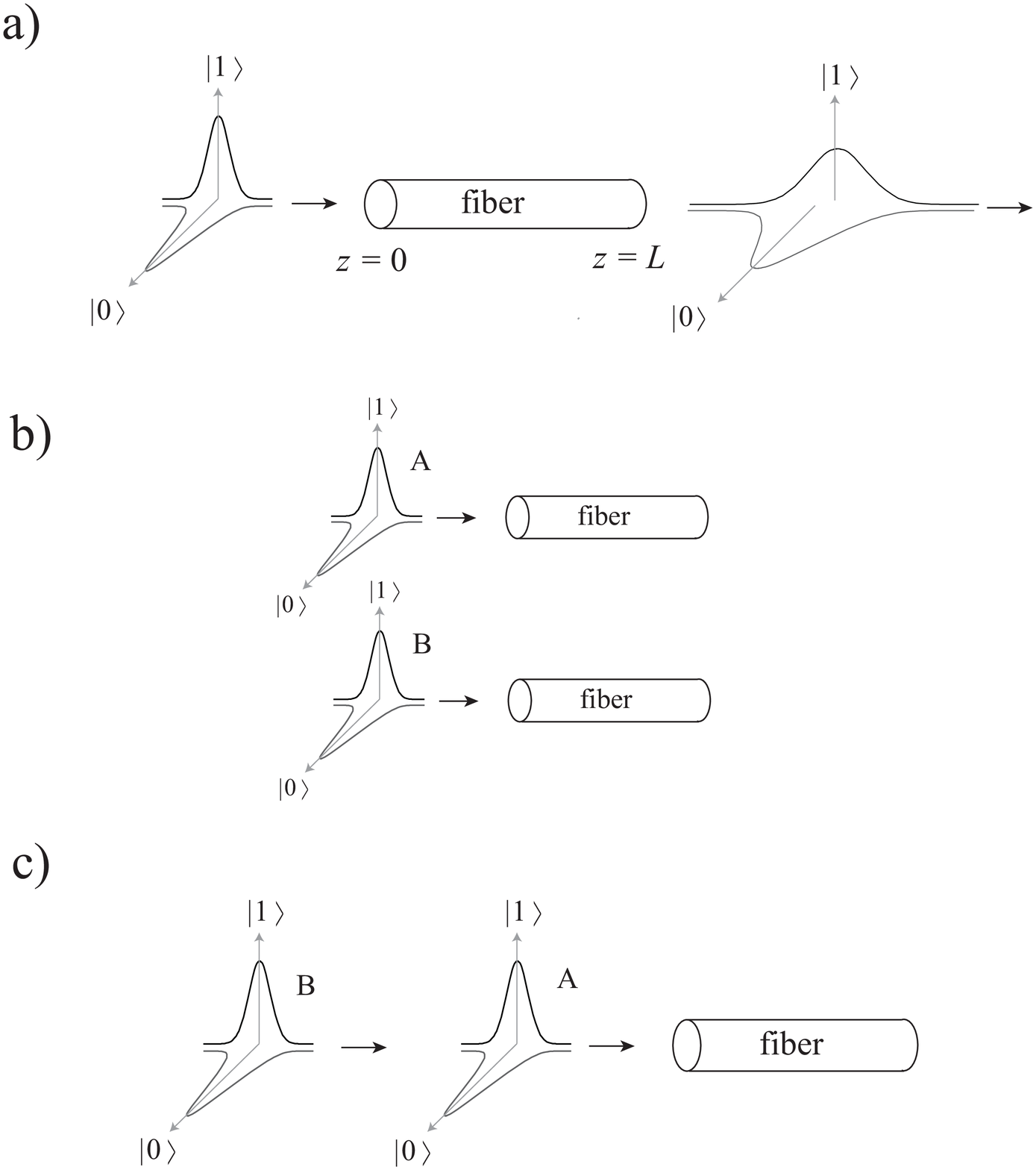}
\caption{\label{fig:figfiber1} Single photon pulses passing through
fibers of length $L$ along $z$ direction: a) A single-photon input
pulse; b) two-photon entangled input pulses each passing through
separate fibers; and c) two-photon entangled input pulses passing
through a common fiber.}
\end{figure}

\section{A Model for Stochastic Polarization Mode Dispersion}
In this section, we present basic equations for an optical pulse in
optical fibers subjected to stochastic PMD \cite{PMDstat}. In a
single mode fiber, only two modes with orthogonal polarizations are
supported. For an optical pulse propagating along $z$ in a linear
and birefringent medium, fiber birefringence ${\bf
b}(\omega,z)=b_1{\bf e}_1+b_2{\bf e}_2+b_3{\bf e}_3$, where ${\bf
e}_j$ are unit vectors in the Stoke's space, alters the polarization
the field depending on its frequencies. The birefringence effect on
the field modes, each of frequency $\omega$, can be derived directly
from the wave equation, which gives
\begin{equation}
\frac{\partial^2}{\partial z^2}{\bf E}(\omega,z)+{\bf k}^2(\omega,z)
{\bf E}(\omega,z)=0.
\end{equation}
Here ${\bf k}^2(\omega,z)\equiv \beta_0^2(\omega,z) +
\beta_0(\omega,z) {\bf b}(\omega,z)\cdot{\bm \sigma}$ is the
propagation tensor, with $\beta_0(\omega,z)$ as the common
propagation constant, and the vector ${\bm \sigma}$ is formed by the
Pauli matrices given by ${\bm \sigma}=\hat{\sigma}_1{\bf
e_1}+\hat{\sigma}_2{\bf e_2}+\hat{\sigma}_3{\bf e_3}$, with
\begin{equation}
\hat{\sigma}_1=
\left(
\begin{array}{cc}
 0  & 1        \\
 1   & 0
\end{array}
\right), \ \ \
\hat{\sigma}_2=
\left(
\begin{array}{cc}
 0  & -i        \\
 i   & 0
\end{array}
\right), \ \ \
\hat{\sigma}_3=
\left(
\begin{array}{cc}
 1  & 0        \\
 0   & -1
\end{array}
\right).
\end{equation}
Extracting the fast oscillating common phase from the field, we
define ${\bf
E}(\omega,z)=E(\omega)\exp\left[-i\int_0^z\beta_0(\omega,z')
dz'\right]{\bf A}(\omega,z)$ with $E(\omega)$ as the amplitude of
the mode and ${\bf A}(\omega,z)$ as its frequency spectrum of the
2-dimensional Jones vector $\tilde{{\bf A}}(t,z)$ \cite{Agrawal}
\begin{equation}
{\bf A}(\omega,z) = \frac{1}{\sqrt{2\pi}}\int dt \tilde{{\bf
A}}(t,z)\exp(i\omega t).
\end{equation}
We adopt the adiabatic approximation, assuming that the birefringence
vector ${\bf b}(\omega,z)$ and thus the polarization of the pulse
varies slowly along $z$ \cite{PMD_review}. As a consequence the term
$\frac{\partial^2}{\partial z^2}{\bf A}(\omega,z)$ becomes
negligible, and the effects of birefringence on the spectrum ${\bf
A}(\omega,z)$, as the pulse passes through the fiber, can be
described by the following first order differential equation \cite{PMD_review}
\begin{equation}\label{eq:singlediffeq_z}
i\frac{\partial}{\partial z}{\bf A}(\omega,z) =\frac{1}{2}
\left[{\bf b}(\omega,z)\cdot{\bm \sigma}\right] {\bf A}(\omega,z).
\end{equation}
Note that for non-dispersive channels, ${\bf b}(\omega,z)\cdot{\bm
\sigma}$ gives eigenvalues $\pm \omega [n_F(z)-n_S(z)]/c$ determined
by the refractive indices associated with the fast mode $n_F$ and
slow mode $n_S$.

It is important to note that Eq.~(\ref{eq:singlediffeq_z}) is a kind
of Schr\"odinger equation with the `Hamiltonian' $\hat{
h}=\frac{1}{2}\left[{\bf b}(\omega,z)\cdot {\bm \sigma}\right]$, in
the form similar to that of a spin-half system interacting with a
magnetic field. For deterministic evolution, we can express the
output pulse with a unitary transformation,
\begin{equation}\label{eq:onephoton_trans}
{\bf A} (\omega,L)= e^{-i {\hat T} \int^L_0 \hat{{h}} dz} {\bf
A}(\omega,0)
\end{equation}
with ${\hat T}$ referring to a position ordered integration
analogous to the time ordered integration in quantum theory. In
terms of column vectors in Jones space, we have
\begin{equation}\label{eq:onephoton}
{\bf A}(\omega,0)\equiv\left(
\begin{array}{cc}
 C_{1}^{{\rm in}} (\omega)        \\
 C_{0}^{{\rm in}} (\omega)
\end{array}
\right)\ \ \ \ \ \mbox{and}\ \ \ \ \ {\bf A}(\omega,L)\equiv\left(
\begin{array}{cc}
 C_{1}^{{\rm out}} (\omega)        \\
 C_{0}^{{\rm out}} (\omega)
\end{array}
\right),
\end{equation}
where $C_{j}^{{\rm in}}$ and  $C_{j}^{{\rm out}}$  $(j=1,0)$ are
polarization amplitudes for input and output fields obeying the
normalization condition: $|C_{0}^{{\rm in}}|^2+|C_{1}^{{\rm
in}}|^2=|C_{0}^{{\rm out}}|^2+|C_{1}^{{\rm out}}|^2=1$. For
conceptual clarity, we will assume that the regions $z <0$ and $z
>L$ are free of birefringence. In this way,
${\bf A}(\omega,z<0)={\bf A}(\omega,0)$ and ${\bf
A}(\omega,z>L)={\bf A}(\omega,L)$.

The stochastic nature of our PMD model originates from the
randomness of ${\bf b}(\omega,z)$. In this paper we adopt the
assumption from \cite{Agrawal}, that the randomness is due to the
fluctuations of optical axis along the fiber. In addition, the
frequency dependence of ${\bf b}(\omega,z)$ is determined by fiber
material properties only and therefore should not change over
different positions. In this way it is plausible to assume
\cite{Agrawal}:
\begin{equation}\label{eq:sepomegaz}
{\bf b}(\omega,z)=f(\omega){\bf b}(z)
\end{equation}
where ${\bf b}(z)$ is stochastic, and the $f(\omega)$ is a
deterministic function defined by the material and it can be
expressed in terms of a Taylor series about the peak frequency
$\omega_0$ of input pulses:
\begin{equation}
f(\omega)=\gamma \omega +\varsigma\omega(\omega-\omega_0) + \cdots .
\end{equation}
Note that in non-dispersive media, only the first term remains and
most of our discussions below will be based on such an
approximation. To specify the statistics of ${\bf b}(z)$, we assume
that ${\bf b} (z)$ is a random process with zero mean, and it has
the two-point correlation function \cite{Agrawal,PMDstat}:
\begin{equation} \label{eq:biref_gaussian}
\overline {{\bf b}(z_1){\bf b}(z_2)} = {\eta}^2
\overleftrightarrow{{\bf I}} \delta(z_2-z_1).
\end{equation}
Here the bar refers to the ensemble average, and ${\eta}$
characterizes the average strength of birefringence. We remark that
the assumption of delta correlation function
(\ref{eq:biref_gaussian}) is not a strict requirement. As long as
the correlation length of ${\bf b} (z)$ is sufficiently short, in
the sense that ${\bf A} (\omega,z)$ does not change significantly
within the correlation length, then it is justified to employ the
Markovian approximation in obtaining the master equation in later
sections.

\section{General Description of Single and Two-photon states}

Let us first examine a deterministic situation corresponding to a
given realization of ${\bf b}(\omega,z)$. In the previous section,
we have seen that if an incoming wave of  frequency $\omega$
incident from the left with the polarization state $(C_{1}^{{\rm
in}} ,C_{0}^{{\rm in}})$, then according to
Eq.~(\ref{eq:onephoton_trans}) and Eq.~(\ref{eq:onephoton}), there
is an outgoing wave propagating to the right with the polarization
state $(C_{1}^{{\rm out}} ,C_{0}^{{\rm out}})$. Since the system (or
the wave equation) is linear, an incoming single-photon follows the
same transformation rule to become an outgoing photon. Let $\phi
(\omega)$ be the frequency envelope of the input single-photon wave
packet, then the input and output state vectors, denoted by $| {\Psi
_{{\rm in}}^{(1)} } \rangle$ and $| {\Psi _{{\rm out}}^{(1)} }
\rangle$, take the form:
\begin{eqnarray}
&& | {\Psi _{{\rm in}}^{(1)} } \rangle  = \int_{}^{} {d\omega \phi
(\omega )} \left| \omega  \right\rangle  \otimes \left[ {C_{1
}^{{\rm in}}(\omega) \left| 1 \right\rangle  + C_{0 }^{{\rm in}}
(\omega) \left| 0 \right\rangle } \right]
 \\
&& | {\Psi _{ \rm out}^{(1)} } \rangle  = \int_{}^{} {d\omega \phi
(\omega )} \left| \omega  \right\rangle  \otimes \left[ {C_{1
}^{{\rm out}}(\omega) \left| 1 \right\rangle  + C_{0 }^{{\rm out}}
(\omega) \left| 0 \right\rangle } \right]
\end{eqnarray}
where $\left|\omega\right\rangle$ is the frequency basis vector
defined in the birefringence free $[{\bf b}(\omega,z)=0]$ system,
and $|1 \rangle$ and $|0\rangle$ respectively correspond to
horizontal and vertical polarization basis vectors. We point out
that in writing Eq. (10) and (11), we have employed a rotating frame
such that the phase factor $e^{-i\omega t}$ due to the free field
evolution of $| \omega \rangle$ has been removed. This is equivalent
to the representation in interaction picture. If Schr\"odinger
picture is needed, we just need to replace $| \omega \rangle$ by
$e^{-i\omega t_{{\rm in}}} | \omega \rangle$ in Eq. (10), and $|
\omega \rangle$ by $e^{-i\omega t_{{\rm out}}} | \omega \rangle$ in
Eq. (11), with $t_{{\rm in}}$ and $t_{{\rm out}}$ are instant of
times defining the input and output states. Both $t_{{\rm in}}$ and
$t_{{\rm out}}$ should be chosen in such a way that the input and
output wave packets are far away from the birefringence interaction
region.

It is important to note that if we treat the interaction length
$L=z$ (Fig. 1) as a parameter, and let $|\Psi _{ \rm out}^{(1)} (z)
\rangle$ be the output state corresponding to a birefringence fiber
of length $z$, then $|\Psi _{ \rm out}^{(1)} (z) \rangle$ is
governed by the Schr\"odinger-like equation according to Eq.
(\ref{eq:singlediffeq_z}), i.e.,
\begin{equation}\label{eq:singledif}
i\frac{\partial}{\partial z}{| {\Psi _{ \rm out}^{(1)} (z) }
\rangle} = \hat {\cal H} ^{(1)} (z) | {\Psi _{ \rm out}^{(1)} (z)}
\rangle
\end{equation}
where $z$ plays the role of time, and
\begin{equation} \label{eq:singledH}
\hat {\cal H}^{(1)} (z) = \int d\omega
\left|\omega\right\rangle\left\langle\omega\right|\otimes
\frac{1}{2} \left[{\bf b}(\omega,z)\cdot{\bm \sigma}\right]
\end{equation}
plays the role of Hamiltonian.

In the case of two-photon states we will restrict our discussion to
systems involving two distinct single-photon pulses, A and B, such
that each pulse contains a single photon (Fig. 1). In other words,
we can label the photons as two subsystems A and B. The
distinguishability of the two photons can be achieved in two
physical situations of interest here. The first situation is
illustrated in Fig. 1b in which the two single-photon pulses
individually propagate in two different optical fibers, and the
second situation is when two spatially (or temporally) separated
photons propagate in the same fiber (Fig. 1c). In both cases, we
have the input-output state vectors:
\begin{eqnarray}\label{eq:twophoton_in}
&& | {\Psi _{\rm in}^{(2)} } \rangle  = \int_{}^{} {\int_{}^{}
{d\omega_A d\omega_B\phi (\omega_A ,\omega_B)} \left| {\omega_A
,\omega_B} \right\rangle  \otimes \sum\limits_{s_A,s_B = 0,1}^{}
{C_{s_As_B}^{{\rm
in}} } } (\omega_A ,\omega_B)\left| {s_A,s_B} \right\rangle \\
&& | {\Psi _{\rm out}^{(2)} } \rangle  = \int_{}^{} {\int_{}^{}
{d\omega_A d\omega_B \phi (\omega_A ,\omega_B)} \left| {\omega_A
,\omega_B} \right\rangle  \otimes \sum\limits_{s_A,s_B = 0,1}^{}
{C_{s_As_B}^{{\rm out}} } } (\omega_A ,\omega_B)\left| {s_A,s_B}
\right\rangle
\end{eqnarray} with $\phi(\omega_A,\omega_B)$ being the normalized frequency
envelope of the input 2-photon wave packet, i.e., $\int \int
d\omega_A d\omega_B |\phi (\omega_A ,\omega_B)|^2 =1$, and the
$C_{s_As_B}^{{\rm in}}(\omega) $ and $C_{s_As_B}^{{\rm out}}$
describe the joint polarization amplitudes for input and output
states at the corresponding frequencies. We remark that input states
with non-factorizable $\phi(\omega_A,\omega_B)$ correspond to
frequency entangled states, and similarly, non-separable
$C_{s_As_B}^{{\rm in}}$ means polarization entanglement.

The fact that the two distinguishable single-photons do not interact
allows us to treat their evolution by the transformation rule as in
the case of single photon. Similar to Eq.(\ref{eq:singledif}), we
can treat the interaction length $L=z$ as a parameter and obtain the
Schr\"odinger equation,
\begin{equation}\label{eq:twodiff1}
i\frac{\partial}{\partial z}{| {\Psi _{ \rm out}^{(2)} (z) }
\rangle} = \hat {\cal H} ^{(2)} (z) | {\Psi _{ \rm out}^{(2)} (z)}
\rangle
\end{equation}
with
\begin{equation} \label{eq:twoH1}
\hat {\cal H}^{(2)} (z) = \frac{1}{2} \int d\omega \left[{\bf
b}_1(\omega,z)\cdot {\bm \sigma^{(A)}} \otimes
\left|\omega\right\rangle_A\left\langle\omega\right|+{\bf
b}_2(\omega,z)\cdot {\bm \sigma^{(B)}} \otimes
\left|\omega\right\rangle_B\left\langle\omega\right| \right]
\end{equation}
for the case of separate fiber (Fig. 1b), and  ${\bf b}_1(\omega,z)$
and ${\bf b}_2(\omega,z)$ are birefringence vectors of the two
fibers. The ${\bm \sigma^{(A)}}$ and ${\bm \sigma^{(B)}}$ are Pauli
vectors for photons A and B.

In the case of common fiber (Fig. 1c), we have
\begin{equation} \label{eq:twoHcommon}
\hat {\cal H}^{(2)} (z) = \frac{1}{2} \int d\omega {\bf
b}(\omega,z)\cdot \left[ {\bm \sigma^{(A)}} \otimes
\left|\omega\right\rangle_A\left\langle\omega\right|+ {\bm
\sigma^{(B)}} \otimes
\left|\omega\right\rangle_B\left\langle\omega\right| \right].
\end{equation}
which indicates that both photons experience the same birefringence
interaction. Note that the pure state vectors discussed above can be
considered as quantum trajectories, corresponding to a single
realization of a birefringence vector. To address the stochastic
problem, we will need to to perform averaging via the density
matrices. In later sections, theoretical analysis of the density
matrices will be carried out according to the master equations for
each of the cases in Fig. 1.

\section{Decoherence of a single-photon state}
Let us first discuss the master equation describing a single-photon
state passing through the fiber. From Eq.~(\ref{eq:singledif}), the
density operator $\hat{\rho}^{(1)}_s(z)$  of the state obeys the
equation
\begin{eqnarray}
i\frac{\partial \hat{\rho}^{(1)}_s(z) }{\partial z} = {[\hat{{\cal
H}}^{(1)} (z),\hat{\rho}^{(1)}_s(z)]},
\end{eqnarray}
where the subscript $s$ stands for a given realization of
birefringence defining $\hat{{\cal H}}^{(1)} (z)$, i.e., before
taking the ensemble average. We follow the standard strategy to
obtain the master equation \cite{bloch},
\begin{eqnarray}\label{eq:single_master}
\frac{\partial}{\partial z}\hat{\rho}^{(1)}=&&\frac{\eta^2}{4}
\sum_{i=1}^3\left[2\hat{\Gamma}_i \hat{\rho}^{(1)} \hat{\Gamma}_i
  -\hat{\Gamma}_i\hat{\Gamma}_i\hat{\rho}^{(1)}-\hat{\rho}^{(1)}\hat{\Gamma}_i\hat{\Gamma}_i\right],
\end{eqnarray}
where we have used $\hat{\rho}^{(1)} \equiv
\overline{\hat{\rho}^{(1)}_s}$ in order to simplify the notation,
and $\hat{\Gamma}_i\equiv \int d\omega  f(\omega)\left|\omega
\right\rangle\left\langle\omega\right|\otimes \hat{\sigma_i}$ is
defined. For later purposes the single-photon density matrix can be
expressed explicitly by,
\begin{eqnarray}\label{eq:single_density}
\hat{\rho}^{(1)} (z) ={\int_{}^{} {\int_{}^{} {d\omega d\omega '
\sum\limits_{s,s' = 0,1}^{} \rho _{ss'} \left( {\omega ,\omega '; z}
\right)} } } \left| {\omega ,s} \right\rangle \left\langle {\omega
',s'} \right|
\end{eqnarray}
where $\rho _{ss'} \left( {\omega ,\omega '; L} \right)$ are matrix
elements.

We point out that the derivation of the master equation
Eq.~(\ref{eq:single_master}) is based on the
Bloch-Redfield-Wangsness approach known in nuclear magnetic
resonance literature \cite{bloch}. Alternatively, the same master
equation can be derived by treating the fiber medium as a bath with
many degrees of freedom \cite{gardiner,limitations}. The former
approach, which we adopt here, can be understood more transparently
by considering the fiber as composed of concatenating uncorrelated
short sections of length $\Delta z$, and $\Delta z$ is set to be
long compared with the coherence length of ${\bf b} (z)$, but is
small so that the change of state can be approximated by keeping the
Dyson series up to the second order in $\hat{{\cal H}}^{(1)}$. Then
by Markovian approximation Eq.~(\ref{eq:biref_gaussian}), the master
equation (\ref{eq:single_master}) in fact corresponds to the `coarse
rate of variation' $\Delta \hat{\rho}^{(1)}_s / \Delta z$ upon
ensemble average.

Now, we consider a general single-photon pulse which is initially
polarized along $\left| 1\right\rangle$,
\begin{eqnarray}\label{eq:single_initial}
\left|\Psi_{in}\right\rangle=\left|1\right\rangle\otimes\int d\omega
\phi(\omega) \left|\omega\right\rangle.
\end{eqnarray}
We remark that the result is the same for arbitrary polarization
direction due to the symmetry caused by the randomizing effect of
the birefringence fluctuations, and hence we set it to $\left|
1\right\rangle$ for convenience. The input pulse envelope is set as
a Gaussian wave packet
\begin{eqnarray}\label{eq:single_Ainitial}
\phi(\omega)=\left(\frac{2}{\kappa^2\pi}\right)^{1/4}\exp\left[\frac{-(\omega-\omega_0)^2}{\kappa^2}\right],
\end{eqnarray}
where $\kappa$ indicates the width of the Gaussian envelope and the
peak frequency $\omega_0$ is in the optical range. With the input
condition $\hat{\rho}_{in}=\left|1\right\rangle\left\langle
1\right|\otimes\int\int d\omega d\omega'
\phi(\omega)\phi^*(\omega')\left|\omega\right\rangle\left\langle\omega'\right|$,
we solve the output state governed by the master equation
Eq.~(\ref{eq:single_master}), and the details are presented in
Appendix (\ref{app1}). The solution for the density matrix is given
by,
\begin{eqnarray}\label{eq:single_diff2}
\rho_{11}(\omega,\omega';L) &=&\frac{1}{2}\phi(\omega)\phi^*(\omega')\left(e^{-\lambda_1 L}+e^{-\lambda_2 L}\right)\nonumber\\
\rho_{00}(\omega,\omega';L)&=&\frac{1}{2}\phi(\omega)\phi^*(\omega')\left(e^{-\lambda_1 L}-e^{-\lambda_2 L}\right),
\end{eqnarray}
where the values of $\lambda_i \geq 0 $ at given $\omega$, $\omega'$ are
\begin{eqnarray}\label{eq:lambda}
\lambda_1(\omega,\omega')&= & \frac{3\eta^2}{4}[f(\omega)-f(\omega')]^2 \nonumber\\
\lambda_2(\omega,\omega')&= & \frac{\eta^2}{4}[3f(\omega)^2+3f(\omega')^2+2f(\omega)f(\omega')],
\end{eqnarray}
It is interesting to note that $\lambda_1$ is a difference of the
frequency profiles and $\lambda_2$ is a sum, giving $\lambda_1 \ll
\lambda_2$ in the optical region. The off diagonal elements are
$\rho_{10}(\omega,\omega';L) = \rho_{01}(\omega,\omega';L)=0$.
Having solved the matrix elements, the output density matrix
$\hat{\rho}^{(1)}(L)$ can therefore be obtained according to
Eq.~(\ref{eq:single_density}). It can be observed that in the long
length limit $L\rightarrow \infty$, a complete depolarization
occurs, since only diagonal elements
$\rho_{11}(\omega,\omega;L)=\rho_{00}(\omega,\omega;L)=\frac{1}{2} |\phi(\omega)|^2$ remain.

\subsection{Pulse spreading}
Let us introduce the quantized field operator,
\begin{equation}
\hat{{\cal E}}(z,t)=\int d\omega \ e^{i\omega (z-ct)/c}
\hat{a}_\omega + h.c.
\end{equation}
where $\hat{a}_\omega$ is the corresponding annihilation operator.
To visualize the change of output pulse shape due to PMD, we examine
the intensity operator $\hat{I} \equiv \hat{\cal E}^\dag \hat{\cal
E} $ by evaluating its expectation value with respect to the output
pulse. The calculation is quite tedious, and we present the results
only. For linear non-dispersive fibers $f(\omega)=\gamma \omega$, we
find that the spatial dependence of the averaged output intensity
associated with the two polarization states are given by
\begin{eqnarray}\label{eq:single_diff3}
\langle \hat{I}(\tau) \rangle_1 &=&\sqrt{\frac{\pi}{2}}\left\{\frac{e^{-\frac{\kappa^2\tau^2}{2[1+6L/(L_c\nu^2)]}}}{\sqrt{1+6L/(L_c\nu^2)}}+\frac{e^{-\frac{\kappa^2\tau^2}{2[1+2L/(L_c\nu^2)]}-\frac{8L/L_c}{1+4L/(L_c\nu^2)}}}{\sqrt{[1+2L/(L_c\nu^2)][1+4L/(L_c\nu^2)]}}\right\} \nonumber \\
\langle \hat{I}(\tau)\rangle _0
&=&\sqrt{\frac{\pi}{2}}\left\{\frac{e^{-\frac{\kappa^2
\tau^2}{2[1+6L/(L_c\nu^2)]}}}{\sqrt{1+6L/(L_c\nu^2)}}-\frac{e^{-\frac{\kappa^2\tau^2}{2[1+2L/(L_c\nu^2)]}-\frac{8L/L_c}{1+4L/(L_c\nu^2)}}}{\sqrt{[1+2L/(L_c\nu^2)][1+4L/(L_c\nu^2)]}}\right\}
\end{eqnarray}
where $\tau= z/c-t$, with $t$ being a sufficiently long time so that
the entire pulse has exited from the birefringence fiber of length
$L$. In addition we have defined the characteristic decoherence
length
\begin{equation}
L_c\equiv \left(\frac{\eta^2}{4}\gamma^2\omega_0^2\right)^{-1},
\end{equation}
and the dimensionless ratio $\nu\equiv \omega_0/\kappa$.

In the narrow bandwidth case where $\nu\gg 1$, the second term in
Eq. (\ref{eq:single_diff3}) decay approximately exponentially with
the decay length $L_c/8$. Such a decaying length scale is much
shorter than that in the first term. Therefore as $L\gg L_c$
increases, only the first term remains, equalizing both elements and
hence showing depolarization. It can be seen that the width of the
output pulse is approximately $c\sqrt{2[1+6L/(L_c\nu^2)]}/\kappa$,
which has a $\sqrt{L}$ dependence when $L \gg L_c\nu^2$ is
sufficiently large. Such a $\sqrt{L}$ dependence were also reported
in general classical consideration \cite{Gisin1995}. The pulse width
at the output is greater than that of input, thus sets a lower bound
to the distance between input pulses in order to allow the output
pulses to be non-overlapping, or distinguishable.

\subsection{Purity functions}
Next, we investigate the decoherence of the single-photon pulse
through its purity defined as $\mu\equiv {\rm Tr}(\hat{\rho}^2)\leq
1$. Purity is closely related to the reciprocal number of effective
modes required to contain the whole state, hence the equality sign
holds only when the state is pure, i.e. can be completely
represented by one mode of the basis. We calculate the frequency
purity, the polarization purity and the overall purity.

To obtain the frequency purity, we trace the polarization freedom of
$\hat{\rho}^{(1)}(L)$ and consider only the frequency DOF. From
Eq.~(\ref{eq:single_density}) and Eq.~(\ref{eq:single_diff2}),
noting that
\begin{eqnarray}
{\rm Tr}_{s}(\hat{\rho}^{(1)}(L))=\int\int d\omega d\omega' \phi(\omega)\phi^*(\omega')e^{-\lambda_1 L} \left|\omega\right\rangle \left\langle\omega'\right|,
\end{eqnarray}
the frequency purity is given by,
\begin{eqnarray}
\mu_{\omega}(L)&=&{\rm Tr}\left\{[{\rm Tr}_{s}(\hat{\rho}^{(1)}(L))]^2\right\}=\frac{1}{\sqrt{1+6L/(L_c\nu^2)}},
\end{eqnarray}
which decays algebraically with $L$. Note that $\mu_{\omega}(L)$
depends on a decay length scale $\nu^2 L_c$, meaning that
$\mu_{\omega}(L)$ has a slower decay rate for a greater $\nu$, i.e.,
narrower spectrum. A remark is that the effective number of
frequency modes (as measured by $\mu_{\omega}^{-1}$) required to
represent the state increases from $1$ to infinity as $L$ increases.
In fact, $\mu_{\omega}^{-1}$ shares a similar functional form with
the output pulse width described in the previous section. This
suggests that the pulse widening can be a measure of PMD decoherence
of frequency variables for Gaussian initial states.

Next we find the polarization purity of $\hat{\rho}^{(1)}(L)$ after
tracing the frequency freedom from $\hat{\rho}^{(1)}(L)$. Noting
that
\begin{eqnarray}
{\rm Tr}_{\omega}(\hat{\rho}^{(1)}(L))&=&{\int_{}^{} {d\omega [\rho _{11} \left( {\omega ,\omega;
L} \right)\left| {1} \right\rangle \left\langle
{1} \right|+\rho _{00} \left( {\omega ,\omega;
L} \right)} }  \left| {0} \right\rangle \left\langle {0} \right|],
\end{eqnarray}
the polarization purity $\mu_s(L)$ is therefore
\begin{eqnarray}\label{eq:single_totalpur}
\mu_s(L)&=&{\rm Tr}\left\{[{\rm Tr}_{\omega}(\hat{\rho}^{(1)}(L))]^2\right\}\nonumber\\
&=&\left[\int d\omega \rho _{11} \left( \omega ,\omega;
L \right)\right]^2+\left[\int d\omega \rho _{00} \left( \omega ,\omega;
L \right)  \right]^2\nonumber\\
&=&\frac{1}{2}\left[1+\frac{1}{1+4L/(L_c\nu^2)}e^{-\frac{16L/L_c}{1+4L/(L_c\nu^2)}}\right],
\end{eqnarray}
decreasing from $1$ to the minimum value $1/2$, meaning that the
state is more spread out in the 2-dimensional Jones space to the
fully mixed situation as fiber length increases. Note that the decay
length scale for $\mu_s(L)$ is $L_c/16$ in the narrow bandwidth case
with $\nu \gg 1$, which is shorter than that for $\mu_{\omega}(L)$.
In addition, appearance of the exponential factor in
Eq.~(\ref{eq:single_totalpur}) indicates a faster decay rate than
that of $\mu_{\omega}(L)$.

The total purity is also found by
\begin{eqnarray}
\mu_{total}(L)&=&{\rm Tr}\left\{[\hat{\rho}^{(1)}(L)]^2\right\}\nonumber\\
&=&\frac{1}{2}\left\{\frac{1}{\sqrt{1+6L/(L_c\nu^2)}}+\frac{e^{-\frac{16L/L_c}{1+4L/(L_c\nu^2)}}}{\sqrt{[1+2L/(L_c\nu^2)][1+4L/(L_c\nu^2)]}}\right\},
\end{eqnarray}
again having a slower rate of decay for a greater $\nu$, and obeying
$\mu_{\omega}(L)\cdot\mu_s(L)\leq\mu_{total}(L)$. We conclude that
decoherence for single-photon state in Eq.~(\ref{eq:single_initial})
is complete only upon $L\rightarrow\infty$.

\section{Disentanglement of two-photon states in separate fibers}
In this section we study the disentanglement of an entangled
two-photon state propagating along separate fibers, each with random
birefringence ${\bf b}_i(z)$ for $(i=1,2)$. We assume that they are
made from the same material, hence obeying $\left\langle{\bf
b}_i(z_1){\bf b}_j(z_2)\right\rangle = \eta^2
\overleftrightarrow{{\bf I}} \delta_{ij}\delta (z_2-z_1)$. The
`Hamiltonian' in this case is given by Eq. (\ref{eq:twoH1}). With
similar assumptions as in the single-photon situation, we obtain the
master equation governing the ensemble averaged density matrix
$\hat{\rho}^{(2)} \equiv \overline {\hat{\rho}^{(2)}_s}$
\begin{eqnarray}\label{eq:two_sep_master}
\frac{\partial}{\partial z}\hat{\rho}^{(2)}=&&\frac{\eta^2}{4} \sum_{j=A,B}\sum_{i=1}^3 \left[2\hat{\Gamma}^{(j)}_i\hat{\rho}^{(2)} \hat{\Gamma}^{(j)}_i-\hat{\Gamma}^{(j)}_i\hat{\Gamma}^{(j)}_i\hat{\rho}^{(2)}-\hat{\rho}^{(2)}\hat{\Gamma}^{(j)}_i\hat{\Gamma}^{(j)}_i\right],
\end{eqnarray}
where $\hat{\Gamma}^{(j)}_i\equiv \int d\omega
f(\omega)\left|\omega\right\rangle_j\left\langle\omega
\right|\otimes \hat{\sigma}_i^{(j)}$ for the two photons $j=A,B$. It
can be noted that the master equation consists of two decoupled
parts, each for an individual fiber. As in the previous section, it
will be convenient to write the two-photon density matrix explicitly
\begin{eqnarray}
\hat{\rho}^{(2)} (z) = {\int_{}^{} {\int_{}^{} {\int_{}^{}
{\int_{}^{} {} } d\omega _A d\omega _B d\omega _A 'd\omega _B '
\sum\limits_{s_A s_B s_A' s_B' = 0,1}^{} \rho _{s_A
 s_B s_A' s_B'} \left( {\omega _A ,\omega _B ,\omega _A
',\omega _B '; z}
\right)} } } \nonumber \\
\times \left| {\omega _A ,\omega _B } \right\rangle \left\langle
{\omega _A ',\omega _B '} \right| \otimes \left| {s_A ,s_B }
\right\rangle \left\langle {s_A',s_B'} \right|.
\end{eqnarray}
with the matrix elements $\rho _{s_A
 s_B s_A' s_B'} \left( {\omega _A ,\omega _B ,\omega _A
',\omega _B '; z} \right)$.

In this section we investigate an initially hyperentangled state
\cite{hyperent}, a polarization Bell state with frequency
entanglement:
\begin{eqnarray} \label{eq:two_sep_initial}
| {\Psi _{\rm in}^{(2)} } \rangle=\int\int \phi(\omega_A,\omega_B) d\omega_A d\omega_B \left|\omega_A,\omega_B\right\rangle \otimes \left|\psi_{\rm Bell}\right\rangle,
\end{eqnarray}
where the frequency envelope is assumed to be a double Gaussian,
with a peak frequency $\omega_0$, as follows:
\begin{eqnarray} \label{eq:two_freqenv}
\phi(\omega_A,\omega_B)=\sqrt{\frac{4\alpha\beta}{\pi}}\exp[-\alpha^2(\omega_A-\omega_B)^2 -\beta^2(\omega_A+\omega_B-2\omega_0)^2],
\end{eqnarray}
where the width $\alpha>\beta$ corresponds to a more frequency
correlated state and $\beta>\alpha$ indicates a more frequency
anticorrelated state. We consider an input singlet pulse
$\left|\psi_{\rm
Bell}\right\rangle=\frac{1}{\sqrt{2}}(|10\rangle-|01\rangle)$. The
evolution of the four Bell states, including the singlet state and
the triplet states, follow same calculation steps and thus only the
singlet state evolution is discussed in detail, presented in
Appendix (\ref{app:two_sep}).

Following the master equation Eq.~(\ref{eq:two_sep_master}), the
evolution of the only non-zero density matrix elements at
$\omega_A,\omega_B,\omega_A',\omega_B'$ can be found as
\begin{eqnarray}\label{eq:two_sep_diff2}
\rho_{1111}&=&\rho_{0000}=\frac{1}{4}\phi(\omega_A,\omega_B)\phi^*(\omega_A',\omega_B')\left(e^{-\zeta_1 L}-e^{-\zeta_4 L}\right)\nonumber\\
\rho_{1010}&=&\rho_{0101}=\frac{1}{4}\phi(\omega_A,\omega_B)\phi^*(\omega_A',\omega_B')\left(e^{-\zeta_1 L}+e^{-\zeta_4 L}\right)\nonumber\\
\rho_{1001}&=&\rho_{0110}=-\frac{1}{2}\phi(\omega_A,\omega_B)\phi^*(\omega_A',\omega_B')e^{-\zeta_4 L},
\end{eqnarray}
where the values of $\zeta_i$ at given $\omega_A$, $\omega_B$, $\omega_A'$ and $\omega_B'$ are shown in Eq.~(\ref{eq:two_sep_lambda}). We can see that $\zeta_1 \ll \zeta_4$ in the optical region. We remark that the Bell states have zero projection to the spaces characterized by $\zeta_2$ and $\zeta_3$ and therefore $\zeta_2$ and $\zeta_3$ do not contribute to the state evolution, which is shown in Appendix (\ref{app:two_sep}). In particular, $\zeta_1=0$ only for the diagonal elements, and thus the steady state is a completely depolarized one, with $\rho_{1111}=\rho_{1010}=\rho_{0101}=\rho_{0000}=\frac{1}{4}|\phi(\omega_A,\omega_B)|^2$.

\subsection{Characterization of entanglement in terms of negativity}

According to Peres and Horodecki's PPT (positive partial
transposition) criterion \cite{PPTperes,PPThoro1, PPThoro}, if the
partial transposition of a bipartite density matrix (denoted by
$\rho^{T_A}$) has one or more negative eigenvalues, then the state
is an entangled state. The negativeness of $\rho^{T_A}$ turns out to
be a necessary and sufficient condition of mixed state entanglement
for two-qubit states and bipartite Gaussian states. To quantify how
much entanglement survives by PMD decoherence in polarization and
frequency variables, we calculate the negativity of the
corresponding DOFs.

The negativity ${\cal N}$ of the state  $\hat{\rho}$ is defined by
\cite{monotone}
\begin{equation}
{\cal N} = \frac{||\hat{\rho}^{T_A}||-1}{2}
\end{equation}
which is an entanglement monotone under local operation and
classical communication \cite{monotone}. Specifically, the
polarization negativity is found by first tracing the frequency
variables and taking the trace norm of the partial transposition
($s_A\leftrightarrow s_A'$), i.e.
\begin{equation}
{\cal N}_{s}= \frac{||\{{\rm
Tr}_{\omega_A,\omega_B}[\hat{\rho}^{(2)}(L)]\}^{T_A}||-1}{2}.
\end{equation}
Similarly, the frequency negativity is obtained by finding the trace
norm of the partially transposed density matrix with the
polarization variables traced, i.e.,
\begin{equation}
{\cal N}_{\omega} = \frac{||\{{\rm
Tr}_{s_A,s_B}[\hat{\rho}^{(2)}(L)]\}^{T_A}||-1}{2}.
\end{equation}
Equivalently, negativities can be obtained by summing the absolute
value of negative eigenvalues of the partially transposed matrices
\cite{negsum}, which we adopt in the following sections.

\subsection{Polarization disentanglement}
\begin{figure}
\includegraphics [width=7 cm] {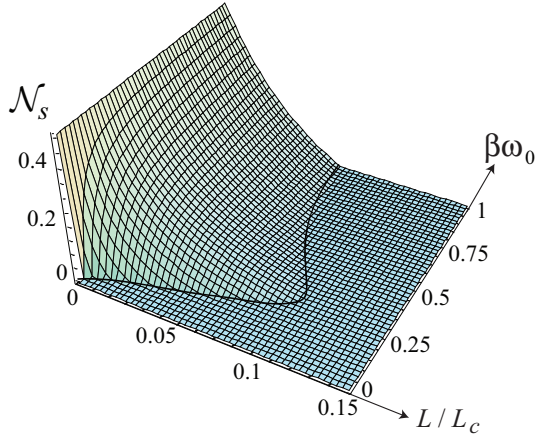}
\caption{\label{fig:two_sep_neg} The polarization negativity for the
two-photon separate fiber case for $\alpha \omega_0 =1000$ with varying
$\beta$ and $L$. The black solid curve indicates the critical
disentanglement length.}
\end{figure}

Now we discuss polarization disentanglement of the two-photon output
state by finding the  negativity ${\cal N}_{s}$ of the state. We
first obtain
\begin{eqnarray}
\{{\rm
Tr}_{\omega_A,\omega_B}[\hat{\rho}^{(2)}(L)]\}^{T_A}&=&\int\int
d\omega_A d\omega_B \left(
\begin{array}{cccc}
 \rho_{1111}&0&0&0       \\
 0&\rho_{1010}&\rho_{1001}&0 \\
 0&\rho_{0110}&\rho_{0101}&0\\
 0&0&0&\rho_{0000}
\end{array}
\right)^{T_A}\nonumber \\
&=&\left(
\begin{array}{cccc}
 \frac{1}{4}(1-\chi)&0&0&-\frac{1}{2}\chi       \\
 0&\frac{1}{4}(1+\chi)&0&0 \\
 0&0&\frac{1}{4}(1+\chi)&0\\
 -\frac{1}{2}\chi&0&0&\frac{1}{4}(1-\chi)
\end{array}
\right),
\end{eqnarray}
where by assuming $f(\omega)=\gamma \omega$,
\begin{eqnarray}\label{eq:two_sep_chi}
 \chi= \exp\left.\left[\frac{-16L/L_c}{1+(2L/\beta^2\omega_0^2L_c)}\right]\right/\sqrt{\left(1+\frac{2L}{\alpha^2\omega_0^2L_c}\right)\left(1+\frac{2L}{\beta^2\omega_0^2L_c}\right)}.
\end{eqnarray}
The polarization negativity ${\cal N}_{s}$ can thus be found as
\begin{eqnarray} \label{eq:two_sep_negpol}
{\cal N}_{s} = \frac{3}{4}\chi-\frac{1}{4}.
\end{eqnarray}
Note that Eq.~(\ref{eq:two_sep_negpol}) is applicable to all four
Bell states since we have the freedom to redefine the polarization
bases and phases in the second fiber due to the symmetry caused by
the  stochastic birefringence fluctuations. We also remark that
${\cal N}_{s}$ has a decay length scale $L_c/16$, and we note that the
same decay length scale exists for polarization purity in the single
photon case. In addition, we observe that finite length
disentanglement is possible when $\chi\leq 1/3$, which can be solved
numerically. As the exponential factor of $\chi$ is depending on the
value of $\beta$, we fix $\alpha$ and plot the trend of polarization
negativity with varying $\beta\omega_0$ in
Fig.~\ref{fig:two_sep_neg}. We see that in general the critical
disentanglement length increases and peaks at some value of
$\beta\omega_0$, then level off to some finite value as
$\beta\omega_0$ tends to infinity.

\subsection{Frequency disentanglement}
To calculate the frequency negativity, we trace the polarization
variables as follows:
\begin{eqnarray}
{\rm Tr}_{s_A,s_B}[\hat{\rho}^{(2)}(L)]&=&\int\int\int\int d\omega_A d\omega_B d\omega_A' d\omega_B'\nonumber \\ &&\phi(\omega_A,\omega_B)\phi^*(\omega_A',\omega_B')e^{-\zeta_1 L}|\omega_A, \omega_B\rangle \langle \omega_A', \omega_B'|
\end{eqnarray}
It is insightful to note that this equation holds not only for Bell
states, but for any general initial polarization states having the
form of Eq.~(\ref{eq:two_sep_initial}) \cite{two_sep_freqneg}. An
important consequence is that frequency entanglement in separate
fibers is independent to initial polarization status. This result is
in contrast to the polarization negativity we presented previously,
which is dependent on the initial frequency envelope shape.

The partial transposition of the frequency density matrix can be
found by $\rho^{T_A}(\omega_A,\omega_B,\omega_A',\omega_B') =\langle
\omega_A',\omega_B|{\rm
Tr}_{s_A,s_B}[\hat{\rho}^{(2)}(L)]|\omega_A,\omega_B'\rangle$. For
linear non-dispersive medium $f(\omega)=\gamma\omega$, the frequency
density matrix remains Gaussian. This allows us to determine its
negativity can be analytically from the general formula for Gaussian
states in \cite{PoonNeg},
\begin{eqnarray}
{\cal N}_{\omega} =\left\{
\begin{array}{cc}
 \frac{1}{2}\left(\sqrt{\frac{\alpha^2\omega_0^2L_c}{\beta^2\omega_0^2L_c+3L}}-1 \right) \ \ \ \ \mbox{for} \ \ \ \ \alpha^2\omega_0^2L_c > \beta^2\omega_0^2L_c+3L      \\
 \frac{1}{2}\left(\sqrt{\frac{\beta^2\omega_0^2L_c}{\alpha^2\omega_0^2L_c+3L}}-1 \right) \ \ \ \ \mbox{for} \ \ \ \ \beta^2\omega_0^2L_c > \alpha^2\omega_0^2L_c+3L
\end{array}
\right. .
\end{eqnarray}
It is interesting to see that finite length frequency disentanglement occurs when
\begin{eqnarray}\label{eq:two_sep_polfin}
L\geq \frac{\omega_0^2}{3}|\alpha^2-\beta^2|L_c,
\end{eqnarray}
which is universal to any initial polarization state. The expression
signifies that the system is `less robust', i.e. having the critical
length of disentanglement tends to zero, if the initial frequency
envelope has $\alpha\approx\beta$. Note that the polarization
entanglement mainly depends on the value of $\beta$, while the
frequency entanglement increases with the difference between
$\alpha^2$ and $\beta^2$.

\section{Disentanglement of two-photon states in a common fiber}
Next, we examine the decoherence problem for two photons propagating
in the same fiber of length $L$. Physically, we can distinguish the
two photons by spatially separate them by a small distance, so small
that the two photons still experience the same stochastic
interactions with the fiber birefringence, yet far apart enough to
be distinguishable, depending on the width of the output pulse which
is discussed quantitatively in the single-photon section. The
`Hamiltonian' in this case is given by Eq.(\ref{eq:twoHcommon}). The
master equation is as follows:
\begin{eqnarray}\label{eq:two_com_master}
\frac{\partial}{\partial z}\hat{\rho}^{(2)}=&&\frac{\eta^2}{4}
\sum_{i=1}^3  \left[2\hat{\Lambda}_i \hat{\rho}^{(2)} \hat{\Lambda}_i-\hat{\Lambda}_i\hat{\Lambda}_i\hat{\rho}^{(2)} -\hat{\rho}^{(2)}\hat{\Lambda}_i\hat{\Lambda}_i \right],
\end{eqnarray}
where
\begin{eqnarray}
\hat{\Lambda}_i &\equiv& \int\int d\omega_A d\omega_B[f(\omega_A) \hat{\sigma}^{(A)}_i+f(\omega_B) \hat{\sigma}^{(B)}_i] \otimes     \left|\omega_A,\omega_B\right\rangle\left\langle\omega_A,\omega_B\right|.
\end{eqnarray}
We remark that the form of this equation is similar to the single
photon master equation Eq.~(\ref{eq:single_master}), but with the
collective operator $\hat{\Lambda}_i$.

It is known that two photons of same frequencies in singlet
polarization states lie in DFS, and their resistance to decoherence
has been verified experimentally \cite{singletexpt}. The reason
behind the decoherence free effect is that they are eigenstates of
$\hat{{\cal H}}^{(2)}$ with an zero eigenvalue, meaning that these
states do not evolve as they propagate. The key is the identical
coupling with the environment for both of the photons
\cite{limitations}. In practice, however, photons generally have
fluctuations in frequencies variables. In the case of down
conversion systems, the two photons are in fact anti-correlated due
to energy conservation. It is thus of interest to examine the
robustness of entanglement for singlet polarization states with a
general Gaussian frequency envelope $\phi(\omega_A,\omega_B)$,
\begin{eqnarray}\label{two_com_singlet}
| {\Psi _{\rm in}^{(2)} } \rangle=\int\int \phi(\omega_A,\omega_B)
e^{-i \omega_A z_0 /c} d\omega_A d\omega_B
\left|\omega_A,\omega_B\right\rangle
\otimes\left[\frac{1}{\sqrt{2}}(\left|10\right\rangle -
\left|01\right\rangle)\right]
\end{eqnarray}
where $\phi(\omega_A,\omega_B)$ is a double Gaussian given in
Eq.~(\ref{eq:two_freqenv}), and the phase factor $e^{-i \omega_A
z_0/c}$ is added in order to displace the peak position of the
photon wave packet A by a distance $z_0$ relative to B. The choice
of separation $z_0$ should be large compared with the width of the
wave packets so that the two photons can be treated as distinct
subsystems, but small enough so that both photons would experience
the same stochastic birefringence. Note that the displacement of
photon A is simply achieved by a local unitary transformation
operator: $\exp(-i z_0 \int \frac{\omega }{c} |\omega \rangle_A
\langle \omega | d \omega )$, and it commutes with $\hat {\cal
H}^{(2)}$ in Eq. (\ref{eq:twoHcommon}). Therefore the phase factor
do not affect the entanglement. For convenience, we will simply
absorb the phase factor into the definition of $|\omega \rangle_A$
in the later calculations.

Following the master equation in Eq.~(\ref{eq:two_com_master}), the
singlet state evolution at $\omega_A,\omega_B,\omega_A',\omega_B'$
are found by the steps in Appendix (\ref{app:two_com}), and are
presented as follows.
\begin{eqnarray}\label{eq:two_com_diff4}
\left(
\begin{array}{cccc}
 \rho_{1111}&\rho_{1110}&\rho_{1101}&\rho_{1100}\\
 \rho_{1011}&\rho_{1010}&\rho_{1001}&\rho_{1000}\\
\rho_{0111}&\rho_{0110}&\rho_{0101}&\rho_{0100}\\
\rho_{0011}&\rho_{0010}&\rho_{0001}&\rho_{0000}
\end{array}
\right)_{L}
&=&
\left(
\begin{array}{cccc}
 \frac{1}{\sqrt{3}}v_2&0&0&0\\
 0&\frac{1}{2}v_1+\frac{1}{2\sqrt{3}}v_2&-\frac{1}{2}v_1+\frac{1}{2\sqrt{3}}v_2&0\\
 0&-\frac{1}{2}v_1+\frac{1}{2\sqrt{3}}v_2&\frac{1}{2}v_1+\frac{1}{2\sqrt{3}}v_2&0\\
 0&0&0&\frac{1}{\sqrt{3}}v_2
\end{array}
\right),
\end{eqnarray}
where
\begin{eqnarray}\label{eq:two_com_vi}
v_1&=&\frac{1}{2}\phi(\omega_A,\omega_B)\phi^*(\omega_A',\omega_B')\left\{\left[\frac{-V^{(1)}_{11}+V^{(1)}_{22}}{\sqrt{\left(V^{(1)}_{11}-V^{(1)}_{22}\right)^2+4{V^{(1)}_{12}}^2}}+1\right]e^{-\xi_1 L} \right. \nonumber\\ &&\ \ \ \ \ \ \ \ \ \ \ \ \ \ \ \ \ \ \ \ \ \ \ \ \ \ \ \ \ \left. +\left[\frac{V^{(1)}_{11}-V^{(1)}_{22}}{\sqrt{\left(V^{(1)}_{11}-V^{(1)}_{22}\right)^2+4{V^{(1)}_{12}}^2}}+1\right]e^{-\xi_2 L} \right\}\nonumber\\
v_2&=&\frac{V^{(1)}_{12}\phi(\omega_A,\omega_B)\phi^*(\omega_A',\omega_B')}{\sqrt{\left(V^{(1)}_{11}-V^{(1)}_{22}\right)^2+4{V^{(1)}_{12}}^2}} \left(-e^{-\xi_1 L} + e^{-\xi_2 L}\right)
\end{eqnarray}
having $\xi_i \geq0$ denoted in Eq.~(\ref{eq:two_com_xi}), with $\xi_1 \gg\xi_2$ in the optical region.

An important situation to notice is when $\xi_2=0$, where no decay
occurs and the matrix elements can survive to infinite fiber length.
It can be noted that $\xi_2=0$ only for three cases, when
$\omega_A=\omega_A'$ and $\omega_B=\omega_B'$, $\omega_A=\omega_B$
and $\omega_A'=\omega_B'$, or $\omega_A=\omega_B'$ and
$\omega_A'=\omega_B$ (Appendix C). Thus we obtain the output state
in the $L\rightarrow \infty$ limit, with the only nonzero terms as
follows:
\begin{eqnarray}\label{eq:two_com_long}
\omega_A=\omega_A',\omega_B=\omega_B': &&\left(
\begin{array}{cccccc}
 \rho_{1111}& \rho_{1010}& \rho_{0101}& \rho_{0000}& \rho_{1001}& \rho_{0110}
\end{array}
\right)_{L\rightarrow \infty}\nonumber\\
&=&\frac{1}{4}\phi(\omega_A,\omega_B)\phi^*(\omega_A,\omega_B)\left(
\begin{array}{cccccc}
 1& 1& 1& 1& 0& 0
\end{array}
\right)\nonumber\\
\omega_A=\omega_B, \omega_A'=\omega_B': &&\left(
\begin{array}{cccccc}
 \rho_{1111}& \rho_{1010}& \rho_{0101}& \rho_{0000}& \rho_{1001}& \rho_{0110}
\end{array}
\right)_{L\rightarrow \infty}\nonumber\\
&=&\frac{1}{2}\phi(\omega_A,\omega_A)\phi^*(\omega_A',\omega_A')\left(
\begin{array}{cccccc}
 0& 1& 1& 0& -1& -1
\end{array}\right)\nonumber\\
\omega_A=\omega_B', \omega_A'=\omega_B:&&\left(
\begin{array}{cccccc}
 \rho_{1111}& \rho_{1010}& \rho_{0101}& \rho_{0000}& \rho_{1001}& \rho_{0110}
\end{array}
\right)_{L\rightarrow \infty}\nonumber\\
&=&\frac{1}{4}\phi(\omega_A,\omega_B)\phi^*(\omega_B,\omega_A)\left(
\begin{array}{cccccc}
 -1& 0& 0& -1& -1& -1
\end{array}\right)
\end{eqnarray}
We remark that the first case refers to a completely depolarized
situation in which the density matrix contains only the diagonal
elements. Furthermore, the second case refers to the decoherence
free situation with both photons having the same frequencies, and
therefore experiencing collective decoherence. In this case it does
not decay and remains as a singlet state for any fiber length $L$.

\subsection{Polarization disentanglement of the singlet state}
Now we discuss polarization disentanglement of the two-photon output
state by finding the corresponding negativity ${\cal N}$ of the
state. We first obtain,
\begin{eqnarray}
\{{\rm Tr}_{\omega_A,\omega_B}[\hat{\rho}^{(2)}(L)]\}^{T_A}=\left(
\begin{array}{cccc}
 \frac{1}{4}(1-\upsilon)&0&0&-\frac{1}{2}\upsilon       \\
 0&\frac{1}{4}(1+\upsilon)&0&0 \\
 0&0&\frac{1}{4}(1+\upsilon)&0\\
 -\frac{1}{2}\upsilon&0&0&\frac{1}{4}(1-\upsilon)
\end{array}
\right),
\end{eqnarray}
where by assuming $f(\omega)=\gamma \omega$,
\begin{eqnarray}
\upsilon &=& \int\int d\omega_A d\omega_B |\phi(\omega_A,\omega_B)|^2 \exp\left\{-2\eta^2[f(\omega_A)-f(\omega_B)]^2L\right\}\nonumber\\
&=&1\left/\sqrt{1+\frac{4L}{\alpha^2\omega_0^2L_c}}\right. .
\end{eqnarray}
The polarization negativity is therefore
\begin{eqnarray}
{\cal N}_{s} = \frac{3}{4}\upsilon-\frac{1}{4}.
\end{eqnarray}
Finite length disentanglement occurs when
\begin{eqnarray}
L\geq 2\alpha^2\omega_0^2L_c.
\end{eqnarray}
We remark that polarization disentanglement is `more robust' if the
two photons have more correlated frequencies, i.e., a larger value
of $\alpha$. In the limit $\alpha\rightarrow\infty$, negativity
never decay, which is the original DFS case where both photons
experience collective decoherence. In addition, comparing with the
polarization negativity of common fiber with that of separate fibers
as in Eq.~(\ref{eq:two_sep_negpol}), we see that the latter has in
general a greater rate of decay due to the exponential decay factor
in Eq.~(\ref{eq:two_sep_chi}).

\subsection{Frequency entanglement of the singlet state}
To investigate frequency entanglement, we first trace the
polarization variables of the output density in
Eq.~(\ref{eq:two_com_diff4}), and take the partial transposition
$\omega_A \leftrightarrow \omega_A'$, obtaining the density matrix
elements as follows,
\begin{equation}\label{eq:two_com_freq}
\rho^{T_A}_{\omega}(\omega_A,\omega_B,\omega_A',\omega_B') =\langle \omega_A',\omega_B|{\rm Tr}_{s_A,s_B}[\hat{\rho}^{(2)}(L)]|\omega_A,\omega_B'\rangle = v_1+\sqrt{3}v_2.
\end{equation}
With a non-Gaussian form Eq.~(\ref{eq:two_com_freq}), there is not a
generally agreed analytical measure for mixed-state entanglement. We
thus attempt to detect the presence of entanglement by evaluating
the positivity of the partially transposed density matrix in
Eq.~(\ref{eq:two_com_freq}). We test its positivity by noting that
for the whole matrix to be positive, any $2\times 2$ submatrices
composed by extracting $4$ points from two of the rows and columns
of the density matrices should be positive. Here we will extract the
points where $\xi_2= 0$ so that they do not decay and are
significant upon the long length limit. For the frequency correlated
case we select  $\omega_A=\omega_B$ and $\omega_A'=\omega_B'$, the
second case of the long length limit in Eq.~(\ref{eq:two_com_long}),
giving the partial transposed elements
$\rho^{T_A}_{\omega}(\omega_B,\omega_A,\omega_A,\omega_B)$, we
consider the $2\times 2$ submatrices for $\omega_A<\omega_B$:
\begin{eqnarray}\label{eq:two_com_22sub}
&&\left(
\begin{array}{cc}
 \rho^{T_A}_{\omega}(\omega_A,\omega_B,\omega_A,\omega_B) & \rho^{T_A}_{\omega}(\omega_A,\omega_B,\omega_B,\omega_A)\\
 \rho^{T_A}_{\omega}(\omega_B,\omega_A,\omega_A,\omega_B) & \rho^{T_A}_{\omega}(\omega_B,\omega_A,\omega_B,\omega_A)
 \end{array}\right)
\end{eqnarray}
Checking the positivity of the matrix, we employ the fact that the
submatrix is negative if and only if the product of off-diagonal
terms is greater than the product of the diagonal terms. Noting that
\begin{equation}
\frac{\rho^{T_A}_{\omega}(\omega_A,\omega_B,\omega_A,\omega_B) \rho^{T_A}_{\omega}(\omega_B,\omega_A,\omega_B,\omega_A)} {\rho^{T_A}_{\omega}(\omega_A,\omega_B,\omega_B,\omega_A)\rho^{T_A}_{\omega}(\omega_B,\omega_A,\omega_A,\omega_B)}=\exp[-4(\alpha^2-\beta^2)(\omega_A-\omega_B)^2],
\end{equation}
we see that all these submatrices are negative when $\alpha>\beta$,
independent of the birefringence interaction length $L$. This
indicates that for singlet polarization states, frequency
entanglement exists at any finite distance $L$ if the frequencies of
the two photons are initially correlated in the form of a double
Gaussian.

On the other hand, if the two photons's frequencies are initially
anticorrelated, i.e., $\beta>\alpha$, we select $\omega_A=\omega_B'$
and $\omega_B=\omega_A'$ for the submatrix positivity test.  This is
guided by third case in Eq.~(\ref{eq:two_com_long}). We therefore
consider the $2\times 2$ submatrices
\begin{eqnarray}\label{eq:two_com_22subanti}
&&\left(
\begin{array}{cc}
 \rho^{T_A}_{\omega}(\omega_A,\omega_A,\omega_A,\omega_A) & \rho^{T_A}_{\omega}(\omega_A,\omega_A,\omega_B,\omega_B)\\
 \rho^{T_A}_{\omega}(\omega_B,\omega_B,\omega_A,\omega_A) & \rho^{T_A}_{\omega}(\omega_B,\omega_B,\omega_B,\omega_B)
 \end{array}\right),
\end{eqnarray}
which are negative if
\begin{eqnarray}\label{eq:two_com_negcon}
\beta > \sqrt{\alpha^2+g(L)},
\end{eqnarray}
where
\begin{eqnarray}\label{eq:two_com_gL}
g(L)\equiv \frac{4L}{\omega_0^2 L_c}-\frac{1}{4(\omega_A-\omega_B)^2}\ln\left[\frac{1}{4}\left(-3+e^{8(\omega_A-\omega_B)^2L/\omega_0^2 L_c}\right)\right] > 0,
\end{eqnarray}
is defined. Particularly in the long length limit, Eq.
(\ref{eq:two_com_negcon}) reduces to
$\beta>\sqrt{\alpha^2+\frac{1}{4(\omega_A-\omega_B)^2}\ln 4}$,
meaning that entanglement persists at long distance if $\beta$ is
sufficiently large.


\section{Conclusion}
To summarize, we discuss quantum disentanglement of frequency and
polarization variables, for photons propagating through fibers with
stochastic PMD. Observing the analogy between the wave propagation
inside the fiber and the Schr\"odinger equation in quantum theory,
master equation method is adopted to analytically solve for the
field density matrix. In this paper we investigate the single-photon
and two-photon cases. For the single photon case, purity function
for each of the DOFs is analytically calculated, quantitatively
determining the degree of mixing, which reveals that complete
decoherence is possible only for infinite fiber length. Pulse width
of the output pulse is also evaluated, determining the minimum
separation of pulses for them to be distinguishable at the output.
Next, for entangled two-photon states with each photon propagating
through a separate fiber, we show that entanglement associated with
frequency and polarization variables can be completely destroyed
after distinct finite propagation length scales. Specifically, for
the hyperentangled state Eq.~(\ref{eq:two_sep_initial}), condition of polarization
disentanglement is found as $\chi < 1/3$ where $\chi$ is defined in
Eq. (\ref {eq:two_sep_chi}), and the condition of frequency
disentanglement is given by Eq.~(\ref{eq:two_sep_polfin}). An interesting fact that
frequency disentanglement does not depend on the initial
polarization status is also revealed. For a singlet polarization
state propagating through a common fiber, we show that polarization
disentanglement in finite length is possible, though having a much
longer critical length of disentanglement than the separate fiber
case. We also consider the frequency entanglement in common fiber,
observing its dependence on the initial frequency envelope. On one
hand, for the frequency correlated parts of the density matrix,
entanglement persists if it already exists at the input, explainable
by the DFS. On the other hand, entanglement can manifest in the
anti-correlated parts of the density matrix if initially the
frequency envelope is sufficiently anticorrelated.

\begin{acknowledgments}
This work is supported by the Research Grants Council of the Hong
Kong SAR, China (Project No. 401307).
\end{acknowledgments}

\appendix

\section{Solution of Master Equation for single-photon state}\label{app1}
We outline the solution of the master equation
Eq.~(\ref{eq:single_master}) in the following. From
Eq.~(\ref{eq:single_master}), the matrix elements
$\rho_{11}(\omega,\omega';z)$ and $\rho_{00}(\omega,\omega';z)$ are
coupled by
\begin{eqnarray}\label{eq:single_diff1}
\frac{\partial}{\partial z}
\left(
\begin{array}{cc}
 \rho_{11}       \\
 \rho_{00}
\end{array}
\right)
=
{\bf M}_1\left(
\begin{array}{cc}
 \rho_{11}       \\
 \rho_{00}
\end{array}
\right)
\end{eqnarray}
where
\begin{eqnarray}
{\bf M}_1=\frac{\eta^2}{4}\left(
\begin{array}{cc}
 -3[f(\omega)^2+f(\omega')^2]+2f(\omega)f(\omega') & 4f(\omega)f(\omega') \\
 4f(\omega)f(\omega') & -3[f(\omega)^2+f(\omega')^2]+2f(\omega)f(\omega')
\end{array}
\right).
\end{eqnarray}
Diagonalizing ${\bf M}_1$, its eigenvalues are $-\lambda_i \leq 0$
as defined in Eq.~(\ref{eq:lambda}), and hence we can find the
solutions by evaluating
\begin{eqnarray}
\left(
\begin{array}{cc}
 \rho_{11}        \\
 \rho_{00}
\end{array}
\right)_{z}
=
\exp({\bf M}_1 z)\left(
\begin{array}{cc}
 \rho_{11}        \\
 \rho_{00}
\end{array}
\right)_{z=0}.
\end{eqnarray}
On the other hand, the off diagonal elements $\rho_{10}(\omega,\omega';z)=\rho_{01}^*(\omega,\omega';z)$ decay individually with the same form
\begin{eqnarray}
\frac{\partial}{\partial z} \rho_{10}(\omega,\omega';z)=-\lambda_2 \rho_{10}(\omega,\omega';z)
\end{eqnarray}
Putting in the input conditions $\rho_{11}(\omega,\omega';0) =\phi(\omega)\phi^*(\omega')$, $\rho_{00}(\omega,\omega';0) =\rho_{10}(\omega,\omega';0) =\rho_{01}(\omega,\omega';0) =0$, we get the solutions in Eq.~(\ref{eq:single_diff2}).

\section{Solution of Master Equation for two-photon state in separate fibers}\label{app:two_sep}
Following the master equation Eq.~(\ref{eq:two_sep_master}), we
found that the matrix elements
$\rho_{1111}(\omega_A,\omega_B,\omega_A',\omega_B';z)$,
$\rho_{1010}(\omega_A,\omega_B,\omega_A',\omega_B';z)$,
$\rho_{0101}(\omega_A,\omega_B,\omega_A',\omega_B';z)$ and
$\rho_{0000}(\omega_A,\omega_B,\omega_A',\omega_B';z)$ are coupled
as follows:
\begin{eqnarray}\label{eq:two_sep_diff1}
\frac{\partial}{\partial z}
\left(
\begin{array}{cccc}
 \rho_{1111}        \\
 \rho_{1010} \\
 \rho_{0101}\\
 \rho_{0000}
\end{array}
\right)
=
{\bf M}_2\left(
\begin{array}{cccc}
 \rho_{1111}        \\
 \rho_{1010} \\
 \rho_{0101}\\
 \rho_{0000}
\end{array}
\right)
\end{eqnarray}
where
\begin{eqnarray}
{\bf M}_2=\frac{\eta^2}{4}\left(
\begin{array}{cccc}
 m_{1} & m_{2} &m_{3}&0 \\
 m_{2}& m_{1} & 0 &m_{3} \\
 m_{3}& 0& m_{1}&m_{2} \\
 0 &m_{3}& m_{2}& m_{1}
\end{array}
\right),
\end{eqnarray}
with the elements
\begin{eqnarray}
m_{1}&= & -3[f(\omega_A)^2+f(\omega_B)^2+f(\omega_A')^2+f(\omega_B')^2]+2f(\omega_A)f(\omega_A') +2f(\omega_B)f(\omega_B')\nonumber\\
m_{2}&= & 4f(\omega_B)f(\omega_B')\nonumber\\
m_{3} &= & 4f(\omega_A)f(\omega_A').
\end{eqnarray}
Diagonalizing ${\bf M}_2$, its eigenvalues are $-\zeta_i \leq 0$ where
\begin{eqnarray}\label{eq:two_sep_lambda}
\zeta_1 &=& \frac{\eta^2}{4}\{3[(f(\omega_A)-f(\omega_A')]^2 +[3(f(\omega_B)-f(\omega_B')]^2\}\nonumber\\
\zeta_2 &=& \frac{\eta^2}{4}\{3[(f(\omega_A)-f(\omega_A')]^2+[3f(\omega_B)^2+2f(\omega_B)f(\omega_B')+3f(\omega_B')^2]\}\nonumber\\
\zeta_3 &=& \frac{\eta^2}{4} \{[3f(\omega_A)^2+2f(\omega_A)f(\omega_A')+3f(\omega_A')^2]+3[(f(\omega_B)-f(\omega_B')]^2\}\nonumber\\
\zeta_4 &=& \frac{\eta^2}{4}\{[3f(\omega_A)^2+2f(\omega_A)f(\omega_A')+3f(\omega_A')^2]\nonumber\\&&\ \ \ \ \ +[3f(\omega_B)^2+2f(\omega_B)f(\omega_B')+3f(\omega_B')^2]\},
\end{eqnarray}
with $\zeta_1$ having the smallest value and $\zeta_4$ the largest.
The corresponding eigenvectors $\varphi_i$ are
\begin{eqnarray}\label{eq:two_sep_eigenvec}
\varphi_1=\frac{1}{2}\left(
\begin{array}{cccc}
 1\\
 1\\
 1\\
 1
\end{array}
\right) ;\varphi_2=\frac{1}{2}\left(
\begin{array}{cccc}
 -1\\
 1\\
 -1\\
 1
\end{array}
\right) ;\varphi_3=\frac{1}{2}\left(
\begin{array}{cccc}
 -1\\
 -1\\
 1\\
 1
\end{array}
\right);\varphi_4=\frac{1}{2}\left(
\begin{array}{cccc}
 1\\
 -1\\
 -1\\
 1
\end{array}
\right)
\end{eqnarray}
Hence, we can find the solutions by evaluating
\begin{eqnarray}
\left(
\begin{array}{cccc}
 \rho_{1111}\\
 \rho_{1010}\\
 \rho_{0101}\\
 \rho_{0000}
\end{array}
\right)_{z}=
\exp({\bf M}_2 z)\left(
\begin{array}{cccc}
 \rho_{1111}\\
 \rho_{1010}\\
 \rho_{0101}\\
 \rho_{0000}
\end{array}
\right)_{z=0}.
\end{eqnarray}
From the eigenvectors in Eq.~(\ref{eq:two_sep_eigenvec}), we see
that the four initial Bell states  have zero projection to the
subspace spanned by $\zeta_2$ and $\zeta_3$ and thus their evolution
are not dependent on these two parameters. In addition, the elements
$\rho_{1001}(\omega_A,\omega_B,\omega_A',\omega_B';z)=\rho_{0110}^*(\omega_A,\omega_B,\omega_A',\omega_B';z)$
and
$\rho_{1100}(\omega_A,\omega_B,\omega_A',\omega_B';z)=\rho_{0011}^*(\omega_A,\omega_B,\omega_A',\omega_B';z)$
decay by themselves with the fastest decay $\zeta_4$:
\begin{eqnarray}
\frac{\partial}{\partial z} \rho_{1001}(\omega_A,\omega_B,\omega_A',\omega_B';z) = -\zeta_4\rho_{1001}(\omega_A,\omega_B,\omega_A',\omega_B';z)
\end{eqnarray}
Putting in the input conditions for the singlet states, i.e.
\begin{eqnarray}
\rho_{1010}(\omega_A,\omega_B,\omega_A',\omega_B';0) &=&\rho_{0101}(\omega_A,\omega_B,\omega_A',\omega_B';0)=\frac{1}{2} \phi(\omega_A,\omega_B)\phi^*(\omega_A',\omega_B')\nonumber \\
\rho_{1001}(\omega_A,\omega_B,\omega_A',\omega_B';0) &=&\rho_{0110}(\omega_A,\omega_B,\omega_A',\omega_B';0)=-\frac{1}{2} \phi(\omega_A,\omega_B)\phi^*(\omega_A',\omega_B'),
\end{eqnarray}
we obtain the solutions in Eq.~(\ref{eq:two_sep_diff2}).

\section{Solution of Master Equation for two-photon state in a common fiber}\label{app:two_com}
Solving the master equation Eq.~(\ref{eq:two_com_master}), we find
that six of the elements are coupled as follows.
\begin{eqnarray}\label{eq:two_com_diff1}
\frac{\partial}{\partial z}
\left(
\begin{array}{cccccc}
 \rho_{1111}\\
 \rho_{1010}\\
 \rho_{0101}\\
 \rho_{0000}\\
 \rho_{1001}\\
 \rho_{0110}
\end{array}
\right)
=
{\bf M}'\left(
\begin{array}{cccccc}
 \rho_{1111}\\
 \rho_{1010}\\
 \rho_{0101}\\
 \rho_{0000}\\
 \rho_{1001}\\
 \rho_{0110}
\end{array}
\right)
\end{eqnarray}
where
\begin{eqnarray}
{\bf M}'=\frac{\eta^2}{4}\left(
\begin{array}{cccccc}
 m'_{1} & m'_{4} &m'_{5}&0 & m'_{6} &m'_{7}\\
 m'_{4}& m'_{2} & 0 &m'_{5}& -m'_{9} &-m'_{8}\\
 m'_{5}& 0& m'_{2}&m'_{4} &-m'_{8}& -m'_{9} \\
 0 &m'_{5}& m'_{4}& m'_{1}&m'_{7}& m'_{6}\\
 m'_{6}& -m'_{9}&-m'_{8}&m'_{7}&m'_{3}&0\\
 m'_{7}&-m'_{8}& -m'_{9}&m'_{6}&0&m'_{3}
\end{array}
\right),
\end{eqnarray}
with the elements, having a unit of the reciprocal of length, as follows.
\begin{eqnarray}
m'_{1}&= & -3[f(\omega_A)^2+f(\omega_B)^2+f(\omega_A')^2+f(\omega_B')^2]-2f(\omega_A)f(\omega_B)-2f(\omega_A')f(\omega_B')\nonumber\\&&+2f(\omega_A)f(\omega_A')+2f(\omega_B)f(\omega_B')+2f(\omega_B)f(\omega_A') +2f(\omega_A)f(\omega_B')\nonumber\\
m'_{2}&= & -3[f(\omega_A)^2+f(\omega_B)^2+f(\omega_A')^2+f(\omega_B')^2]+2f(\omega_A)f(\omega_B)+2f(\omega_A')f(\omega_B')\nonumber\\&&+2f(\omega_A)f(\omega_A')+2f(\omega_B)f(\omega_B')-2f(\omega_B)f(\omega_A') -2f(\omega_A)f(\omega_B')\nonumber\\
m'_{3}&= & -3[f(\omega_A)^2+f(\omega_B)^2+f(\omega_A')^2+f(\omega_B')^2]+2f(\omega_A)f(\omega_B)+2f(\omega_A')f(\omega_B')\nonumber\\&&-2f(\omega_A)f(\omega_A')-2f(\omega_B)f(\omega_B')+2f(\omega_B)f(\omega_A') +2f(\omega_A)f(\omega_B')\nonumber\\
m'_{4}&= & 4f(\omega_B)f(\omega_B')\nonumber\\
m'_{5} &= & 4f(\omega_A)f(\omega_A')\nonumber\\
m'_{6}&= & 4f(\omega_B)f(\omega_A')\nonumber\\
m'_{7}&= & 4f(\omega_A)f(\omega_B')\nonumber\\
m'_{8}&= & 4f(\omega_A)f(\omega_B)\nonumber\\
m'_{9}&= & 4f(\omega_A')f(\omega_B').
\end{eqnarray}
To solve the system more conveniently, we introduce the unitary
operator,
\begin{eqnarray}
{\bf U}=\left(
\begin{array}{cccccc}
 0 & \frac{1}{\sqrt{3}} &-\frac{1}{\sqrt{6}}&0 & -\frac{1}{\sqrt{2}} &0\\
 \frac{1}{2}& \frac{1}{2\sqrt{3}} & \frac{1}{\sqrt{6}} &0& 0&-\frac{1}{\sqrt{2}}\\
 \frac{1}{2}& \frac{1}{2\sqrt{3}}& \frac{1}{\sqrt{6}}&0&0& \frac{1}{\sqrt{2}} \\
 0 &\frac{1}{\sqrt{3}}& -\frac{1}{\sqrt{6}}& 0&\frac{1}{\sqrt{2}}& 0\\
 -\frac{1}{2}&\frac{1}{2\sqrt{3}}&\frac{1}{\sqrt{6}}&-\frac{1}{\sqrt{2}}&0&0\\
 -\frac{1}{2}&\frac{1}{2\sqrt{3}}& \frac{1}{\sqrt{6}}&\frac{1}{\sqrt{2}}&0&0
\end{array}
\right),
\end{eqnarray}
which rotates to the frame where ${\bf M}'(\omega_0,\omega_0,\omega_0,\omega_0)$ is diagonal. Applying the transformation to ${\bf M}'(\omega_A,\omega_B,\omega_A',\omega_B')$,
\begin{eqnarray}
{\bf V}\equiv {\bf U}^\dag {\bf M}' {\bf U} =\frac{\eta^2}{4}\left(
\begin{array}{cccccc}
 V^{(1)}_{11} & V^{(1)}_{12}&0&0 & 0&0\\
 V^{(1)}_{12}& V^{(1)}_{22}&0&0 & 0&0\\
 0&0& V^{(2)}_{11}&0&0&0 \\
 0&0& 0& V^{(3)}_{11} & V^{(3)}_{12}& V^{(3)}_{13}\\
 0&0&0&V^{(3)}_{12}& V^{(3)}_{22}& V^{(3)}_{23}\\
 0&0& 0&V^{(3)}_{13} & V^{(3)}_{23}& V^{(3)}_{33}
\end{array}
\right),
\end{eqnarray}
where we can see that the $6\times 6$ matrix is expressed as a direct sum of three smaller submatrices, and the matrix elements at $\omega_A,\omega_B,\omega_A',\omega_B'$ are
\begin{eqnarray}
V^{(1)}_{11}&=&-3\{[(f(\omega_A)-f(\omega_B)]^2+[(f(\omega_A')-f(\omega_B')]^2\}\nonumber\\
V^{(1)}_{12}&=&2\sqrt{3}[(f(\omega_A)-f(\omega_B)][(f(\omega_A')-f(\omega_B')]\nonumber\\
V^{(1)}_{22}&=&-3[f(\omega_A)^2+f(\omega_B)^2+f(\omega_A')^2+f(\omega_B')^2]-2f(\omega_A)f(\omega_B)-2f(\omega_A')f(\omega_B')\nonumber\\ &&+4f(\omega_A)f(\omega_A')+4f(\omega_B)f(\omega_B')+4f(\omega_B)f(\omega_A') +4f(\omega_A)f(\omega_B')\nonumber\\
V^{(2)}_{11}&=&-3[f(\omega_A)^2+f(\omega_B)^2+f(\omega_A')^2+f(\omega_B')^2]-2f(\omega_A)f(\omega_B)-2f(\omega_A')f(\omega_B')\nonumber\\ &&-2f(\omega_A)f(\omega_A')-2f(\omega_B)f(\omega_B')-2f(\omega_B)f(\omega_A') -2f(\omega_A)f(\omega_B')\nonumber\\
V^{(3)}_{11}&=&-3[f(\omega_A)^2+f(\omega_B)^2+f(\omega_A')^2+f(\omega_B')^2]+2f(\omega_A)f(\omega_B)+2f(\omega_A')f(\omega_B')\nonumber\\ &&-2f(\omega_A)f(\omega_A')-2f(\omega_B)f(\omega_B')+2f(\omega_B)f(\omega_A') +2f(\omega_A)f(\omega_B')\nonumber\\
V^{(3)}_{12}&=&4[f(\omega_B)f(\omega_A')-f(\omega_A)f(\omega_B')]\nonumber\\
V^{(3)}_{13}&=&4[f(\omega_A)f(\omega_B)-f(\omega_A')f(\omega_B')]\nonumber\\
V^{(3)}_{22}&=&-3[f(\omega_A)^2+f(\omega_B)^2+f(\omega_A')^2+f(\omega_B')^2]-2f(\omega_A)f(\omega_B)-2f(\omega_A')f(\omega_B')\nonumber\\ &&+2f(\omega_A)f(\omega_A')+2f(\omega_B)f(\omega_B')+2f(\omega_B)f(\omega_A') +2f(\omega_A)f(\omega_B')\nonumber\\
V^{(3)}_{23}&=&4[f(\omega_B)f(\omega_B')-f(\omega_A)f(\omega_A')]\nonumber\\
V^{(3)}_{33}&=&-3[f(\omega_A)^2+f(\omega_B)^2+f(\omega_A')^2+f(\omega_B')^2]+2f(\omega_A)f(\omega_B)+2f(\omega_A')f(\omega_B')\nonumber\\ &&+2f(\omega_A)f(\omega_A')+2f(\omega_B)f(\omega_B')-2f(\omega_B)f(\omega_A') -2f(\omega_A)f(\omega_B')
\end{eqnarray}
Hence we can solve Eq.~(\ref{eq:two_com_diff1}) by considering these three submatrices, i.e.
\begin{eqnarray}\label{eq:two_com_diff2}
\left(
\begin{array}{cccccc}
 \rho_{1111}\\
 \rho_{1010}\\
 \rho_{0101}\\
 \rho_{0000}\\
 \rho_{1001}\\
 \rho_{0110}
\end{array}
\right)_{z}
=
\exp ({\bf M}'L) \left(
\begin{array}{cccccc}
 \rho_{1111}\\
 \rho_{1010}\\
 \rho_{0101}\\
 \rho_{0000}\\
 \rho_{1001}\\
 \rho_{0110}
\end{array}
\right)_{z=0}=
{\bf U}\exp ({\bf V} L){\bf U}^\dag \left(
\begin{array}{cccccc}
 \rho_{1111}\\
 \rho_{1010}\\
 \rho_{0101}\\
 \rho_{0000}\\
 \rho_{1001}\\
 \rho_{0110}
\end{array}
\right)_{z=0}
\end{eqnarray}
If the input is a singlet state as shown in Eq.~(\ref{two_com_singlet}), only the first $2\times 2$ submatrix is involved, having eigenvalues $-\xi_i \leq 0$ as follows:
\begin{eqnarray}\label{eq:two_com_xi}
\xi_1&=&\frac{\eta^2}{8} \left[-V^{(1)}_{11}-V^{(1)}_{22}+\sqrt{\left(V^{(1)}_{11}-V^{(1)}_{22}\right)^2+4{V^{(1)}_{12}}^2}\right]\nonumber \\
\xi_2&=&\frac{\eta^2}{8} \left[-V^{(1)}_{11}-V^{(1)}_{22}-\sqrt{\left(V^{(1)}_{11}-V^{(1)}_{22}\right)^2+4{V^{(1)}_{12}}^2}\right],
\end{eqnarray}
where $\xi_2\ll \xi_1$ since $V^{(1)}_{11},V^{(1)}_{22}\leq 0$. We remark that $-\xi_1$ and $-\xi_2$ are the two eigenvalues with the smallest magnitudes in the matrix ${\bf V}$.
Eq.~(\ref{eq:two_com_diff2}) thus becomes
\begin{eqnarray}\label{eq:two_com_diff3}
\left(
\begin{array}{cccccc}
 \rho_{1111}\\
 \rho_{1010}\\
 \rho_{0101}\\
 \rho_{0000}\\
 \rho_{1001}\\
 \rho_{0110}
\end{array}
\right)_{z}
=
\phi(\omega_A,\omega_B)\phi^*(\omega_A',\omega_B') {\bf U}\exp ({\bf V} L)\left(
\begin{array}{cccccc}
 1\\
 0\\
 0\\
 0\\
 0\\
 0
\end{array}
\right)
=
{\bf U}\left(
\begin{array}{cccccc}
 v_1\\
 v_2\\
 0\\
 0\\
 0\\
 0
\end{array}
\right),
\end{eqnarray}
with $v_i$ as presented in Eq.~(\ref{eq:two_com_vi}). Then, the
output state can be expressed as in Eq.~(\ref{eq:two_com_diff4}).

Finally, we point out that or $\xi_2$ to be equal $0$, the condition
$\left(V_{12}^{(1)}\right)^2 = V_{11}^{(1)}V_{22}^{(1)}$ has to be
satisfied. Considering the function
$\Omega(\omega_A,\omega_B,\omega_A',\omega_B')
=\left(V_{12}^{(1)}\right)^2 - V_{11}^{(1)}V_{22}^{(1)}$ with all
parameters real. In the case when $\omega_A'=\omega_B'$, we have
$\Omega = 0$ only when $\omega_A=\omega_B$. Otherwise when
$\omega_A'\neq \omega_B'$ and by fixing $\omega_A'$ and $\omega_B'$
we consider $\partial \Omega/\partial \omega_A =0$ and $\partial
\Omega/\partial \omega_B =0$, giving maximum $\Omega$ at only 2
pairs of real condition, namely $\omega_A=\omega_A'$ and
$\omega_B=\omega_B'$, and  $\omega_A=\omega_B'$ and
$\omega_A'=\omega_B$, while these two conditions both give
$\Omega=0$.


\begin{references}
\bibitem{two_sep_freqneg} For any initial entangled two-photon states in the
form Eq.~(\ref{eq:two_sep_initial}), the vector formed by the
elements $\rho_{1111}$, $\rho_{1010}$, $\rho_{0101}$ and
$\rho_{0000}$ for each $\omega_A,\omega_B,\omega_A',\omega_B'$ can
be decomposed into the four eigenvectors in
Eq.~(\ref{eq:two_sep_eigenvec}). Evolution over length for each
$\varphi_i$ imposes an exponential factor of $\exp(-\zeta_i)$.
Noting that only $\varphi_1$ is not traceless, tracing the
polarization variables, we obtain
$\phi(\omega_A,\omega_B)\phi^*(\omega_A,\omega_B)\exp(-\zeta_1)$ as
the output.
\end{references}
\end{document}